\documentclass[prb,twocolumn,amsmath,amssymb,superscriptaddress]{revtex4-2}
\usepackage{graphicx}
\usepackage{dcolumn}
\usepackage{textcomp}
\usepackage{amssymb}
\usepackage{amsmath}
\usepackage{color}
\usepackage{footnote}
\usepackage{microtype}
\usepackage{bm}
\usepackage{hyperref}
\usepackage{array}
\makeatletter
\newcommand*{\rom}[1]{\expandafter\@slowromancap\romannumeral #1@}
\makeatother

\newcolumntype{P}[1]{>{\centering\arraybackslash}p{#1}}
\newcolumntype{M}[1]{>{\centering\arraybackslash}m{#1}}
\newcommand{\beq}{\begin{equation}}
\newcommand{\eeq}{\end{equation}}
\newcommand{\beqa}{\begin{eqnarray}}
\newcommand{\eeqa}{\end{eqnarray}}

\usepackage{dcolumn}

\begin{document}
\title{Robust Zeeman-type band splitting in sliding ferroelectrics}
\author{Homayoun Jafari}
\thanks{These authors contributed equally.}
\affiliation{Zernike Institute for Advanced Materials, University of Groningen, Nijenborgh 4, 9747 AG Groningen, The Netherlands}
\author{Evgenii Barts}
\thanks{These authors contributed equally.}
\affiliation{Zernike Institute for Advanced Materials, University of Groningen, Nijenborgh 4, 9747 AG Groningen, The Netherlands}
\author{Przemys\l aw Przybysz}
\thanks{These authors contributed equally.}
\affiliation{Zernike Institute for Advanced Materials, University of Groningen, Nijenborgh 4, 9747 AG Groningen, The Netherlands}
\affiliation{Department of Solid State Physics, Faculty of Physics and Applied Informatics, University of \L \'{o}d\'{z}, Pomorska 149/153, 90-236 \L \'{o}d\'{z}, Poland}
\author{Karma~Tenzin}
\affiliation{Zernike Institute for Advanced Materials, University of Groningen, Nijenborgh 4, 9747 AG Groningen, The Netherlands}
\affiliation{Department of Physical Science, Sherubtse College, Royal University of Bhutan, 42007 Kanglung, Trashigang, Bhutan}
\author{Pawe\l~J. Kowalczyk}
\affiliation{Department of Solid State Physics, Faculty of Physics and Applied Informatics, University of \L \'{o}d\'{z}, Pomorska 149/153, 90-236 \L \'{o}d\'{z}, Poland}
\author{Pawe\l~Dabrowski}
\affiliation{Department of Solid State Physics, Faculty of Physics and Applied Informatics, University of \L \'{o}d\'{z}, Pomorska 149/153, 90-236 \L \'{o}d\'{z}, Poland}
\author{Jagoda S\l awi\'{n}ska}
\email[Corresponding author:]{\ jagoda.slawinska@rug.nl}
\affiliation{Zernike Institute for Advanced Materials, University of Groningen, Nijenborgh 4, 9747 AG Groningen, The Netherlands}

\date{\today}

\begin{abstract}
Transition metal dichalcogenides (TMDs) exhibit giant spin-orbit coupling (SOC), and intriguing spin-valley effects, which can be harnessed through proximity in van der Waals (vdW) heterostructures. Remarkably, in hexagonal monolayers, the Zeeman-type band splitting of valence bands, which originate from the prismatic crystal field, can reach values of several hundreds of meV, offering significant potential for both fundamental and applied research. While this effect is suppressed in the commonly studied hexagonal (H)-stacked bilayers due to the presence of inversion symmetry, the recent discovery of sliding ferroelectricity in rhombohedral (R-)stacked MX$_2$ bilayers (M=Mo, W; X=S, Se) suggest that the Zeeman effect could be present in these non-centrosymmetric configurations, making it even more intriguing to investigate how the spin-resolved bands would evolve during the phase transition. Here, we perform density functional theory calculations complemented by symmetry analysis to unveil the evolution of ferroelectricity during sliding and the behavior of Zeeman splitting along the transition path. While the evolution of the out-of-plane component of the electric polarization vector resembles the conventional ferroelectric transition, switching between positive and negative values, we observe significant in-plane components parallel to the sliding direction, reaching their maximum at the intermediate state. Moreover, we demonstrate that the R-stacked bilayers exhibit substantial Zeeman-type band splitting, akin to monolayers, which persists throughout the transition path, being allowed by the lack of inversion symmetry. Further analysis of different stacking configurations generated by sliding along various directions confirms that the Zeeman effect in MX$_2$, primarily arising from the polarity of prismatic ligand coordination of the metal atom, is remarkably robust and completely governs the spin polarization of bands, independently of the sliding direction. This resilience promises to maintain robust spin transport in vdW heterostructures based on MX$_2$ bilayers, opening new opportunities for ferroelectric spintronics.
\end{abstract}
\maketitle


\textit{Introduction.} Two-dimensional (2D) transition metal dichalcogenides (TMDs) exhibit a variety of useful and intriguing properties, such as tunable band gaps, high carrier mobilities, spin-valley locking, and strong light-matter interactions, making them ideal for exploring exotic phenomena in fundamental studies and developing advanced electronic devices with enhanced functionalities. Recently, interfacial (or sliding) ferroelectricity has been proposed in 2D materials \cite{binary_compound_2D_ferroelectrics, Ferr_switch_MoS2}, where the broken inversion symmetry induces an interlayer charge transfer, and the electric polarization can be reversed by laterally sliding the layers by applying a vertical electric field \cite{physics_2D_vdW_ferroelectrics}. This effect was initially observed in hexagonal boron nitride (h-BN) \cite{sliding_vdW_ferroelectrics, engineered_2D_ferro_BN} and has been extended to rhombohedral-stacked (R-stacked) MX$_2$ (M=W, Mo; X=Se, S) homobilayers \cite{interfacial_ferroelectricity}, MoS$_2/$WS$_2$ heterobilayers \cite{ferroelectricity_heterobilayer_TMD} and marginally twisted MoS$_2$ bilayers \cite{ferroelectricity_twisted_2D}. The combination of room-temperature (RT) ferroelectricity \cite{transition_order_vdW_ferroelectric} and very low switching barriers (on the order of meV, \cite{ physics_2D_vdW_ferroelectrics}) further enhances the potential applications of these atomically-thin materials making them suitable for ultra-fast non-volatile memories and optoelectronic devices. While the interplay between spin and polar degrees of freedom could bring additional functionalities \cite{meso, noel, varotto}, the studies of band splitting, spin textures, or spin currents in ferroelectric TMDs are still limited \cite{Rashba_splitting_biMoTe2, homayoun}.

The hexagonal monolayers of MX$_2$ manifest a strong spin-orbit coupling (SOC) and the intriguing Zeeman-type band splitting locked at the $K/K'$ valleys, which originates from the unique prismatic ligand coordination of the metal atom characteristic for TMDs. Depending on the choice of metal, the spin-splitting of the valence band can reach up to several hundreds of meV, resulting in giant effective magnetic fields. Crucially, this band splitting and the associated out-of-plane spin texture can be transferred through proximity to other materials in van der Waals (vdW) heterostructures \cite{switch, jose_hugo_garcia}. For instance, in graphene/TMD bilayers, the proximity-induced Zeeman effect enables propagation of spins through the graphene layer despite its weak intrinsic SOC, while also protecting out-of-plane spins against dephasing \cite{ghiasi, valenzuela, ingla_aynes}. Thus, the Zeeman spin texture can be considered a persistent spin texture, ensuring robust spin transport in systems with broken inversion symmetry \cite{tsymbal_psh, snte}. In contrast to monolayers, in H-stacked bilayers, the Zeeman-type band splitting is suppressed by the presence of global inversion symmetry (IS) \cite{zeeman_effect}. Although recent studies on R-stacked bilayers have revealed several non-centrosymmetric configurations with different symmetries that can be accessed through the lateral sliding mechanism \cite{interfacial_ferroelectricity, MX2_stacking, group_theory_ferrobilayers}, the potential presence of the Zeeman effect in these structures and its evolution upon sliding remains to be explored.


In this paper, we perform density functional theory (DFT) calculations for the family of sliding ferroelectrics MX$_2$ to unveil the link between the crystallographic point group symmetry, band splitting and spin (orbital) texture of electronic states in the different regions of the Brillouin zone (BZ). We start with the analysis of the ground state configuration described by the point group C$_{3v}$ that exhibits a tiny out-of-plane electric polarization $P_{\uparrow}/P_{\downarrow}$ confirmed by the experiments, and we show that regardless of the interplay of spin-orbit and layer coupling, a substantial Zeeman-type band spin is present at the $K/K'$ valleys, similarly to the case of monolayers. We further simulate the lateral sliding between $P_{\uparrow}$ and $P_{\downarrow}$ configurations and reveal that the decrease in the out-of-plane polarization is accompanied by the emergence of the in-plane component ($P_{\rightarrow}$) which is parallel to the sliding direction, as determined by the C$_s$ symmetry of the stackings along the sliding path \cite{group_theory_ferrobilayers}. Notably, $P_{\rightarrow}$ is much larger than the $P_{\uparrow}/P_{\downarrow}$ and achieves the largest value at the mid-point of the transition path, where the out-of-plane component is forbidden by the C$_{2v}$ symmetry. In this intermediate state that is not paraelectric, the Zeeman band splitting survives and extends to nearly the entire BZ. The symmetry analysis reveals that the persistent Zeeman effect dictates the spin texture in all stacking configurations accessible via sliding due to both symmetry of the little point group at the corners of BZ and a strong polarity of the prismatic coordination which cannot be easily suppressed in the absence of IS.

\begin{figure*}[ht]
	\includegraphics[width = 0.99\textwidth]{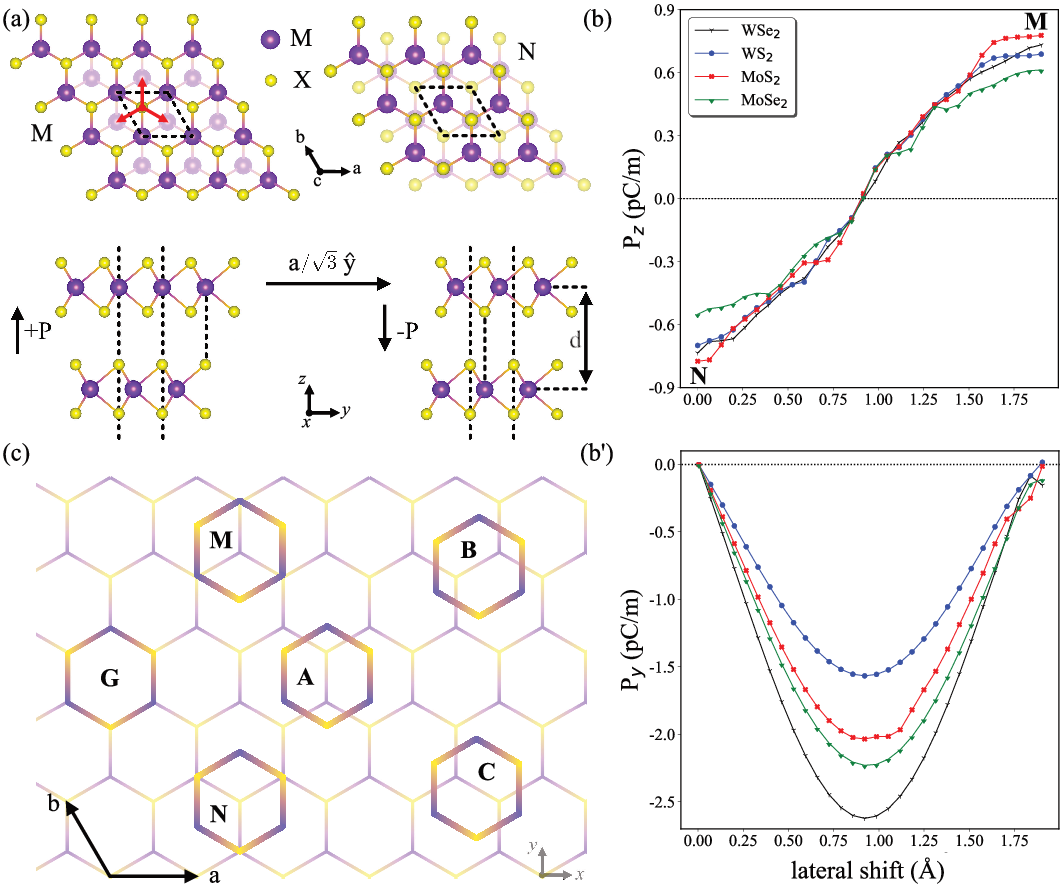}
	\caption{(a) Crystal structure of R-stacked MX$_{2}$ bilayer with polarization upwards ($P_{\uparrow}$) for \textbf{M} stacking and downwards ($P_{\downarrow}$) for \textbf{N} stacking. Top and side views are shown on the top and bottom panels, respectively. Black dashed lines denote the hexagonal unit cell. The red arrows show the three equivalent sliding directions. The interlayer distance $d$ is indicated by the black arrow. (b) Top panel: Out-of-plane electric polarization ($P_z$) versus lateral sliding between \textbf{M} and \textbf{N} configurations calculated for all the considered compounds. The sliding path is along $[120]$ crystallographic direction ($y$ axis). Bottom panel: The same for the in-plane component $P_y$; $P_x = 0$ from symmetry. (c) Schematic view of different stacking configurations. The bottom layer is represented by the semi-transparent honeycomb and the different positions of the top layer are marked by single hexagons. }
	\label{fig:1}
\end{figure*}

\begin{figure*}[ht!]
	\includegraphics[width = 0.99\textwidth]{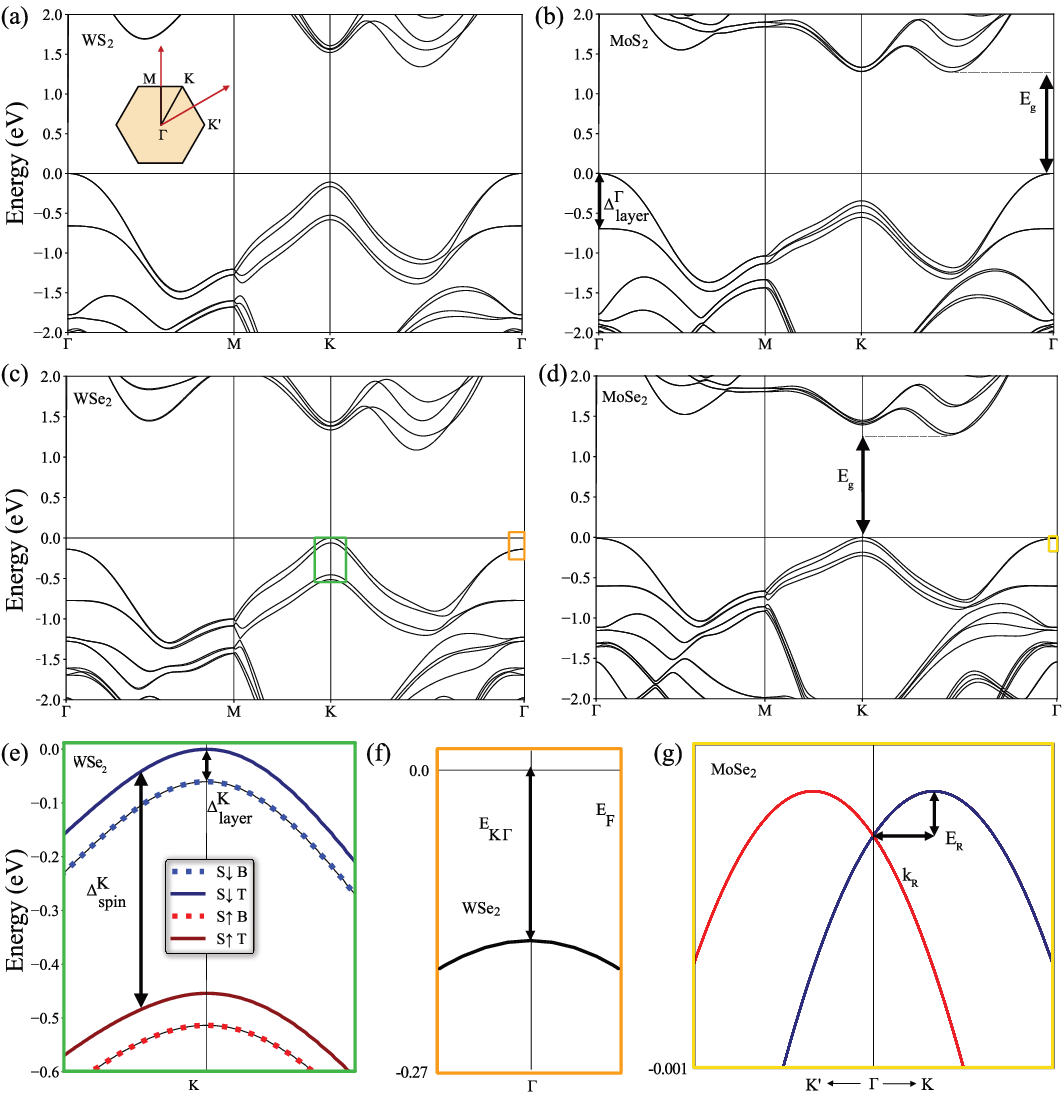}
	\caption{(a)-(d) Band structures of \textbf{M} and \textbf{N} stacking configurations calculated for different MX$_{2}$. The inset shows the Brillouin zone (BZ) and the high-symmetry lines. The arrows indicate the band parameters; $E_{g}$ indirect energy gaps and $\Delta_{\rm layer}^{\Gamma}$ layer splitting at the $\Gamma$ point. (e) The zoom-in of the electronic structure around the $K$ point marked as a green rectangle in panel (c). The coloring denotes out-of-plane spin polarization originating from the Zeeman interaction and the solid/dashed lines represent the bottom and top layers, respectively. The spin-splitting ($\Delta_{\rm spin}^{K}$) and the layer splitting ($\Delta_{\rm layer}^{K}$) parameters are indicated by the arrows. (f) The zoom-in of the electronic structure around the $\Gamma$ point marked as an orange rectangle in panel (c). $E_{K\Gamma}$ denotes the offset between the valence band edge at the $K$ and $\Gamma$ points. (g) The zoom-in of the electronic structure around the $\Gamma$ point marked as a yellow rectangle in panel (d). The colors represent the spin texture perpendicular to the momentum, indicating purely Rashba spin texture. The numerical values of all the parameters are listed in Table I.}
	\label{fig:2}
\end{figure*}

\textit{Structure and ferroelectricity.} TMDs possess multiple crystallographic phases with unique properties determined by the symmetries. Some of these properties also critically depend on the stacking order, such as the evolution of ferroelectricity when adjacent layers are laterally shifted with respect to each other. In contrast to centrosymmetric H-stacked bilayers, the layers in R-stacked configurations are oriented parallel to each other without any relative rotation \cite{stacking_dependent_WSe2, MX2_stacking}. As shown in Fig.\ref{fig:1}(a), the top layer's metal ions align directly above the bottom layer's chalcogen ions and chalcogen ions lie above the centers of the hexagonal rings in the \textbf{M} stacking or vice versa, with chalcogen atoms placed above the metal ions in the \textbf{N} stacking. Both configurations exhibit out-of-plane electric polarization $P_z$, either positive or negative, arising from the electrostatic potential drop between the layers due to the asymmetric charge distribution at the interface \cite{binary_compound_2D_ferroelectrics, interfacial_ferroelectricity}. The ferroelectric transition that corresponds to the switching between \textbf{M} and \textbf{N} stacking orders can be realized by laterally sliding the top layer along one of the three equivalent paths. For our first-principles simulations, we have chosen the path along $[120]$ direction ($y$ axis) and calculated the electric polarization for several intermediate phases between \textbf{M} and \textbf{N} configurations. The values of $P_z$ with respect to the lateral shift are plotted in Fig.\ref{fig:1}(b) for all considered crystals. While the maximal magnitudes are below 1 pC/m, comparable with the previous calculations \cite{binary_compound_2D_ferroelectrics}, MSe$_{2}$ and MS$_{2}$ exhibit the smallest and largest values, respectively (see Table I). The full transition path between the maximal and minimal $P_z$ requires a sliding distance equal to $a/\sqrt3$ that corresponds to approximately 1.8-1.9 \AA. We note that at the mid-point, the out-of-plane polarization vanishes, resembling a paraelectric state.

However, recent research by Ji \textit{et al.} has revealed that the polarization landscape in sliding ferroelectrics is more intricate than previously thought and that in-plane components are allowed by symmetry for most of the parallel stacking configurations \cite{group_theory_ferrobilayers}. Indeed, our calculations have shown in-plane components that emerge along the transition path. Figure \ref{fig:1}(b') shows $P_y$ components calculated for the atomic configurations from the top panel; $P_x$ components have turned out to be negligible. For \textbf{M} and \textbf{N} stacking orders, the values of $P_{y}$ are zero, and they gradually increase when going towards the mid-point \textbf{A}. Remarkably, the in-plane components reach between 1.5 to 2.5 pC/m, significantly exceeding the values of $P_z$. The evolution of $P_y$ is quite similar to the behavior predicted previously for \mbox{h-BN} via DFT and group theory due to the same crystal symmetry. We note, however, that the existence and the possibility of the electric control of these in-plane components in vdW bilayers still need to be verified by experiments.

The allowed direction of $\vec{P}$ can be determined based on the crystallographic point group \cite{group_theory_ferrobilayers} but the crystal symmetries will be also essential for understanding the spin polarization of bands. We will thus start by analyzing the symmetries of the most relevant stacking orders. To provide a clear visual representation of different phases and their point group symmetries, we schematically depict them in Fig.\ref{fig:1}(c). The bottom layer is shown as a semi-transparent honeycomb lattice, while the individual hexagons on top of it correspond to different relative shifts of the top layer. Following the convention from Ref. \cite{group_theory_ferrobilayers}, the fully parallel stacking denoted as \textbf{G} can be characterized by the lateral shift of the top layer relative to the bottom one, parametrized as $(0,0)$ within the basis of the lattice vectors $\vec{a}$ and $\vec{b}$. Similarly, the configurations \textbf{M} and \textbf{N} can be generated by vectors $(\frac{1}{3},\frac{2}{3})$ and $(\frac{2}{3},\frac{1}{3})$, respectively. The stacking orders that correspond to the mid-points along the three equivalent sliding paths are denoted as \textbf{A}, \textbf{B}, and \textbf{C}, and are obtained via shifts by vectors $(\frac{1}{2}, 0)$, $(0, \frac{1}{2})$, and $(\frac{1}{2}, \frac{1}{2})$, respectively.

In each of these phases, specific symmetries prohibit some components of the polarization vector. For example, stacking \textbf{G} is described by the point group D$_{3h}$ generated by the improper rotation S$_{3}$ around the $z$ axis and the mirror plane $m_z$. Similarly to the case of monolayers belonging to the same point group, the electric polarization is entirely prohibited by the presence of improper rotation. The stacking orders \textbf{M} and \textbf{N} possess point group $C_{3v}$ with the three-fold rotation $3_z$ and three mirror planes parallel to the rotation axis. These symmetries only allow for out-of-plane electric polarization, as confirmed by the first principles and experimental studies. The sliding between \textbf{M} and \textbf{N} configurations reduces the symmetry of intermediate phases to C$_s$ with only one mirror plane that is parallel to the sliding direction and perpendicular to the layers. In this case, group theory calculations predict both out-of-plane and in-plane components of electric polarization with the latter parallel to the sliding direction for any of the three equivalent paths \cite{group_theory_ferrobilayers}. It is consistent with our results that show the presence of $P_z$ and $P_y$ components upon sliding along the $y$ axis, as $P_x$ components are forbidden by the mirror reflection $m_y$.  Finally, the mid-points \textbf{A}, \textbf{B}, and \textbf{C} have the symmetry C$_{2v}$, which allows only the in-plane polarization, again in agreement with our calculations.


\begin{table}[ht]
\centering
\caption{Different band parameters defined in Fig.\ref{fig:2} calculated for all considered materials. NG means negligible values that are beyond accuracy of our calculations.
 }
\vspace{0.2cm}
\begin{tabular}{|M{3.0cm}|M{1.2cm}|M{1.2cm}|M{1.2cm}|M{1.2cm}|}
\hline
\vspace*{1cm}
& \textbf{MoS$_2$}  & \textbf{MoSe$_2$} & \textbf{WS$_2$} & \textbf{WSe$_2$}\\
\hline
\vspace*{0.1cm}
$E_{K\Gamma}$,~meV & 344     & 12       & 108       & 138       \\
\hline
\vspace*{0.1cm}
$\lambda_{\Gamma}$,~eV$\cdot$\AA\ & NG     & 0.015       & NG       & 0.012       \\
\hline
\vspace*{0.1cm}
$\Delta_{\rm layer}^{\Gamma}$,~meV  & 693 			& 592 & 659 & 633\\
\hline
\vspace*{0.1cm}
$\Delta_{\rm spin}^{K}$,~meV  		& 148 	                & 184 & 417 & 453 \\
\hline
\vspace*{0.1cm}
$\Delta_{\rm layer}^{K}$,~meV 		& 60 			& 43 & 57 & 60 \\
\hline
\vspace*{0.1cm}
$P_z$,~pC/m 		             		& 0.77 			& 0.59 & 0.69 & 0.73 \\
\hline
\vspace*{0.1cm}
Total band gap,~eV	&1.273	&1.265	&1.341	&1.087 \\
\hline
\end{tabular}
\label{tab:gaps}
\end{table}

\textit{Electronic properties of M- and N-stacked bilayers.} We will start with the discussion of the ground state configurations \textbf{M} (\textbf{N}) with purely out-of-plane electric polarization enforced by the crystallographic point group C$_{3v}$. Based on our first-principles electronic structures summarized in Fig. \ref{fig:2}, all crystals possess indirect energy gaps, with the conduction band minimum (CBM) along the $K-\Gamma$ line, and the valence band maximum (VBM) located either at $\Gamma$ or at $K/K'$ points for the selenides and sulfides, respectively. Apart from the band gaps, the electronic structures are characterized by various parameters, such as band splittings or band-edge energy offsets, defined in Fig.\ref{fig:2}. Their magnitudes calculated for all four crystals are summarized in Table ~\ref{tab:gaps}.

The electronic states show two types of band splitting, one originating from SOC and the other from layer coupling (LC) \cite{ab_texture}. This double splitting is the best visible around the $K/K'$ (Fig.\ref{fig:2} (e)), where the electron states can be described by a minimal Hamiltonian:
\begin{equation}
	\label{eq:Ham}
	H_{K} = - \frac{\Delta_{\rm spin}}{2} \sigma_z - \frac{\Delta_{\rm layer}}{2} \tau_z,
\end{equation}
with ${\sigma}_z$ and ${\tau}_z$ denoting Pauli matrices in the spin and layer space, respectively. The first term is the Dirac mass term, which acts as an effective intrinsic magnetic field, causing Zeeman spin splitting with an energy gap $\Delta_{\rm spin}$. For the $K$ ($K'$) valley, the value of $\Delta_{\rm spin}$ is positive (negative) due to time-reversal symmetry, which flips the spin polarization from the $K$ to $-K$ point. This splitting reaches several hundred of meV (see Table ~\ref{tab:gaps}), similar to monolayers \cite{bhowal, k.p_TMD}. The second term accounts for band energy splitting due to LC. The $\Delta_{\rm layer}$ varies from 50 meV at the $K/K'$ point to 600 meV at the $\Gamma$ point.

To understand the spin-momentum coupling observed at the $\Gamma$ and $K$ points, we conducted a symmetry analysis based on group theory. At the $\Gamma$ point, the symmetry is described by the small group $C_{3v}=\left\{3_z, m_x \right\}$, which consists of a three-fold rotational symmetry axis $3_z$ and a vertical mirror symmetry plane $m_x$, allowing only Rashba-type spin-momentum coupling \cite{zunger}:
\begin{equation}
\label{eq:Rashba}
    \alpha_R \left( k_x \sigma_y - k_y \sigma_x \right),
\end{equation}
By using one-dimensional complex representations of $3_z$, wherein the coordinates $(x,y)$ transform as
\begin{equation}
    3_z \cdot \left(r_+,r_-,z\right) = \left(e^{-i 2\pi/3} r_+,e^{i 2\pi/3} r_-,z\right),
\end{equation}
with $r_\pm = x \pm i y$, we obtain two invariant combinations: $r_+ \sigma_-$ and $r_-\sigma_+$. Their sum leads to Eq.~\eqref{eq:Rashba}, and their difference is a Weyl-type coupling: $\alpha_W \left( k_x \sigma_x + k_y \sigma_y \right)$, which is, however, prohibited by the mirror symmetry $m_x \cdot \left(x,y,z \right) = \left(-x,y,z \right)$. The purely Rashba spin textures are confirmed by our first-principles calculations (see MoSe$_2$ in Fig.\ref{fig:2}(g)), although the corresponding splittings are tiny for all considered materials (see Table ~\ref{tab:gaps}).

\begin{figure*}[ht]
	\includegraphics[width = 0.99\textwidth]{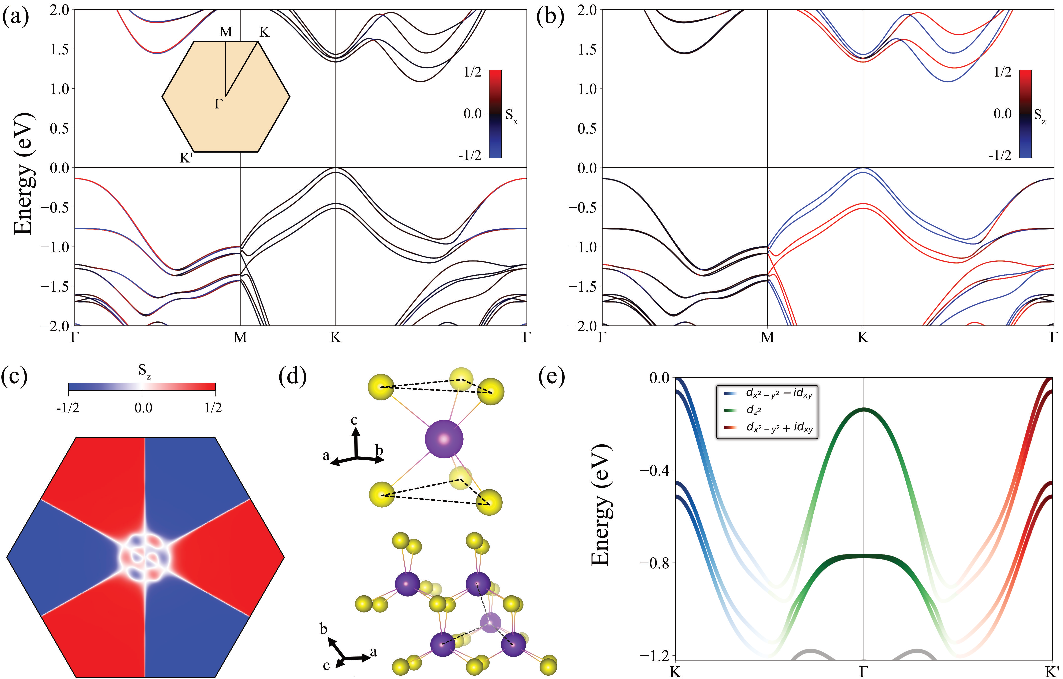}
	\caption{(a)-(b) Spin-resolved band structure of \textbf{N}-stacked WSe$_2$. Panel (a) shows $S_x$ component of the spin texture, $S_y$ is omitted, as it emerges only along $\Gamma-K$ line close to the BZ center. Panel (b) shows the out-of-plane component $S_z$. (c) $S_z$ component calculated for the topmost valence band in the entire BZ. (d)  Illustration of trigonal prismatic ligand coordination in hexagonal MX$_{2}$; top and side views are displayed on top and bottom panels. The dashed lines represent the connection between the ligands and metal atoms located at different layers. (b) Orbital-resolved band structure of WSe$_{2}$ along $K-\Gamma-K'$. }
	\label{fig:3}
\end{figure*}

The spin-momentum coupling at the $K/K'$ points is governed by the small symmetry group $C_3=\left\{ 3_z\right\}$, hence,  both Rashba- and Weyl-type spin textures are allowed. However, our DFT results show that the dominant spin anisotropic interaction in the large $k$-area is the Zeeman-type splitting, which surpasses the Rashba and Weyl interactions and results in a significant band splitting with strongly out-of-plane-polarized spin textures (see Fig.\ref{fig:3} (a)-(c)). It agrees with $3_z$ symmetry which allows for a non-zero spin polarization at the $K$ point \cite{oguchi}, but restricts zero in-plane components for a non-degenerate band, $\psi_K$: $\langle \sigma_{x,y}\rangle=0$, since $\langle{\psi_K}| 3_z^{-1} \sigma_{\pm} 3_z |\psi_{K} \rangle= e^{\pm i 2\pi/3}\langle{\psi_K}| \sigma_{\pm} |\psi_{K}\rangle = 0$. Remarkably, the transition between Zeeman and Rashba spin textures occurs as an abrupt band inversion that interchanges the layer and spin splitting quite close to the $\Gamma$ point (see Fig. \ref{fig:3}(b)), in contrast to a rather smooth decrease of spin splitting along $K-\Gamma$ path found in the monolayer TMDs.

\begin{figure*}[ht]
	\includegraphics[width = 0.99\textwidth]{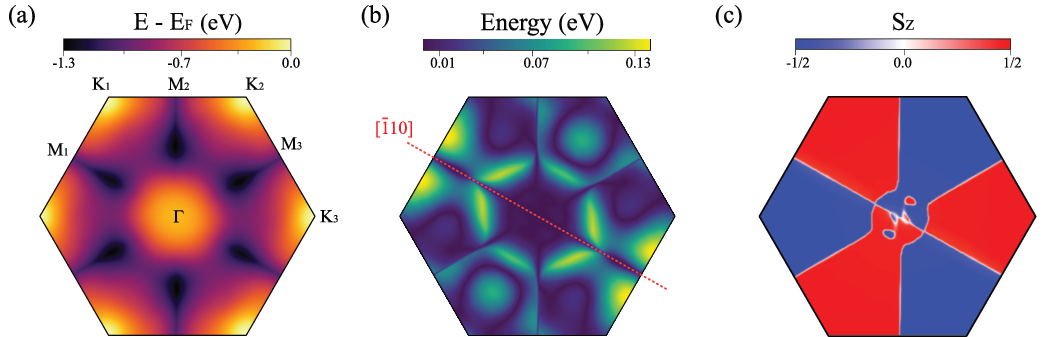}
	\caption{(a) Energy of the topmost valence band calculated over the entire BZ for the intermediate state (\textbf{A}) of WSe$_2$. The bright spots at the corners of the BZ represent the valence band maximum (VBM). The labels indicate the high symmetry points. The energy splitting and spin polarization of this band are shown in panels (b) and (c). The dashed line represents $[\overline{1}10]$ crystallographic direction defined in the reciprocal lattice. In panel (c), only the $S_z$ component of the spin texture is shown, as $S_x$ and $S_y$ are nearly negligible and only present in the close vicinity of $\Gamma$.}
	\label{fig:4}
\end{figure*}

The out-of-plane spin polarization at the $K$ point is permitted by symmetry, but the explanation of the giant splitting requires further analysis. We argue that the large Zeeman interaction comes from a sizable orbital momentum, which is a characteristic feature of strongly polar prismatic ligand ion coordination of transition metals in TMDs, as illustrated schematically in Fig.\ref{fig:3} (d). The prismatic crystal field splits the $d$ orbitals of W (Mo) in three configurations: $\{ d_{xy}, d_{x^2 - y^2}\}$, $\{d_{3z^2 - r^2}\}$  and $\{ d_{xz}, d_{yz}\}$. In all the compounds, the valence band states are constructed from the $d_{xy}$ and  $d_{x^2 - y^2}$ orbitals, specifically
$d_{x^2 - y^2} - i d_{xy} $ and $d_{x^2 - y^2} + i d_{xy}$ at the $K$ and $K'$ point, respectively \cite{bhowal, VBM_in_TMD}. Figure \ref{fig:3} (e) shows the orbital-resolved band structure calculated for WSe$_2$, confirming our analysis. We note that these states are eigenstates of the ${L}_z$ projection of atomic orbital momentum:
\begin{equation}
	\label{eq:L_z}
	{L}_z \left(|{d_{x^2 - y^2}\rangle} \pm i |{d_{xy}\rangle} \right)=
	\pm 2 \left(|{d_{x^2 - y^2}\rangle} \pm i |{d_{xy}\rangle} \right),
\end{equation}
which can be derived using the real space representation of
${L}_z= -i(x\frac{\partial}{\partial y} - y\frac{\partial}{\partial x})$,
where the $|{d_{x^2 - y^2}\rangle} \pm i |{d_{xy}\rangle}$ state transforms as the ${x^2 - y^2} \pm i 2 {xy} $ function. Consequently, the relatively large Zeeman interaction at the $K$ point arises directly from the spin-orbit coupling:
$\lambda \left(\bm L \cdot \bm S \right) = -2\lambda  S_z $, where $S_z =\sigma_z/2$. Here, we used the fact that the expectation values of the $x,y$ orbital momentum components are zero within the manifold of $d_{xy}$ and $d_{x^2 - y^2}$ orbitals, as also dictated by $3_z$ symmetry. Similar arguments hold for monolayer TMDs \cite{zeeman_effect, bhowal}. 
\\



\textit{Electronic properties of the intermediate state.} We will now focus on the electronic structure and spin texture of the intermediate state \textbf{A} shown schematically in Fig.\ref{fig:1}(c). This bilayer possesses mirror symmetry $m_y$ along the sliding direction, a glide mirror $\overline{m}_z$ consisting of a mirror plane parallel to the layers and fractional translation along the $x$ direction, as well as the screw symmetry with the two-fold rotation $2_y$ followed by the translation along the $x$ axis. Although the $C_{2v}$ symmetry results in an orthorhombic structure, we opted to use a hexagonal unit cell for calculations, similar to the \textbf{M} and \textbf{N} stacking configurations. However, it is important to note that the absence of three-fold symmetry in the intermediate state leads to different high-symmetry points in the BZ. Thus, to better visualize the symmetries, we provide maps of the properties of the topmost valence band throughout the full BZ, rather than standard band structures along high symmetry lines. The results of calculations performed for WSe$_2$ are summarized in Fig. \ref{fig:4}(a)-(c).

Figure \ref{fig:4}(a) displays the energy map of the topmost valence band. Although it may initially appear to exhibit three-fold symmetry, a closer examination of the band shape around the BZ center reveals a distortion along the $[\bar{1}11]$ direction in the reciprocal lattice. This anisotropy is even more evident in the map of band splitting shown in Fig. \ref{fig:4}(b), which shows a clear two-fold pattern. We find that different corners of the BZ exhibit distinct electronic structures, in contrast to the \textbf{M} and \textbf{N} configurations. For instance, we observe two different $K$ and $M$ points with the band splitting along $K_1-M_1$ approximately 100 meV larger than along $K_2-M_2$. Surprisingly, the $S_z$ component of the spin texture, plotted in Fig. \ref{fig:4}(c), reveals nearly perfect three-fold symmetry, similar to the $C_{3v}$ configurations (compare with Fig. \ref{fig:3}(c)), except for a tiny region close to $\Gamma$. Notably, the $S_z$ component of the spin texture is now prevalent throughout the entire BZ. This can be rationalized by the mirror symmetry $\overline{m}_z$ that now belongs to the symmetry group of an arbitrary point in the BZ and does not allow an in-plane spin projection. Additionally, linear spin-momentum coupling at the $\Gamma$ point in the BZ is prohibited by $\overline{m}_z$.

\textit{Conclusions.} In summary, we have explored different non-centrosymmetric stacking orders of MX$_2$ bilayers using first-principles calculations and symmetry analysis. We focused first on the behavior of electric polarization upon sliding between two configurations with purely out-of-plane electric dipoles, commonly associated with sliding ferroelectricity. Remarkably, our findings revealed the emergence of in-plane polarization components parallel to the sliding direction, in addition to the perpendicular components, reaching their maximum values at the midpoint of the transition path. The non-zero electric polarization suggests that the intermediate state is neither centrosymmetric nor paraelectric.

Furthermore, the same crystal symmetries that determine the direction of electric polarization also significantly impact the electronic states, particularly the spin texture. In R-stacked bilayers, the band structures exhibit two distinct types of splitting: one originating from SOC and the other from LC. Around the two valleys, the dominant feature is the Zeeman-type band splitting, reaching magnitudes of several hundreds of meV. This remarkable splitting is enabled by the symmetries arising from the prismatic ligand coordination of the metal atoms, which dictate the configuration of the valence states and lead to substantial spin splitting arising directly from SOC, similar to the case of monolayers. Additionally, consistent with time-reversal symmetry, opposite spin moments are observed at the two valleys $K$ and $-K$. An extra splitting of each of these spin-split bands occurs due to LC, but it becomes dominant only in the vicinity of the $\Gamma$ point.

In addition, we have studied spin polarization of the intermediate states along the transition path. The mid-point configuration, characterized by the point group C$_{2v}$, exhibits an enhanced Zeeman-type persistent spin texture that now extends throughout nearly the entire BZ. This behavior is attributed to the presence of glide reflection $\overline{m}_z$, which prevents in-plane spin projection at any point in the reciprocal lattice. While larger regions display the Zeeman-type band splitting, the change from three-fold to two-fold symmetry does not significantly alter the arrangement of spin texture in the BZ. Interestingly, other stacking configurations accessible through sliding, such as those described by point groups C$s$ or C$_2$, exhibit similar Zeeman splitting resembling either the C$_{3v}$ or C$_{2v}$ symmetry configurations. The only way to eliminate this splitting is either by restoring the global inversion symmetry or by twisting the layers to modify the little group of different points in the BZ.

In terms of spin transport, any of the considered bilayers can serve a similar role as monolayers, effectively protecting propagating spins from randomization during scattering. Additionally, by populating one of the valleys, further phenomena like the valley Hall effect can be achieved \cite{mak_science, mak_nano, Pol_in_Cent_TMD}. Breaking time-reversal symmetry can be accomplished through the application of electric or magnetic fields, or via circularly polarized optical excitation \cite{asymmetric_mpi, lin_ahe}. Furthermore, unlike monolayers, the studied structures exhibit sliding ferroelectricity and a complex polarization landscape, offering numerous opportunities for further exploration that include multi-state ferroelectricity and moiré domains in twisted structures \cite{ferroelectricity_twisted_2D, Cumulative_polarization, multi-state}. The combination of these peculiar ferroelectric properties at room temperature, along with robust spin transport, opens up intriguing possibilities for electronic devices endowed with enhanced functionalities.

\begin{table}[ht]
	\centering
	\caption{Structural parameters of all considered crystals; $a$, $d_{\mathrm{ M-X}}$, $d_{\mathrm{ X-X}}$, $\delta_{\mathrm{ X-M-X}}$, $d$ and $t_{\mathrm{ eff}}$ denote lattice constants, bond lengths M-X and X-X, bond angle between X-M-X, interlayer distance, and effective thickness, respectively.}
	\vspace{0.2cm}
	\begin{tabular}{|M{3.0cm}|M{1.2cm}|M{1.2cm}|M{1.2cm}|M{1.2cm}|}
		\hline
		\vspace*{1cm}
		& \textbf{MoS$_2$}  & \textbf{MoSe$_2$} & \textbf{WS$_2$} & \textbf{WSe$_2$}\\
		\hline
		\vspace*{0.1cm}
		$a$ (\AA) \  &3.16 			& 3.28 & 3.15 &3.28\\
		\hline
		\vspace*{0.1cm}
		$d_{\mathrm{ M-X}}$ (\AA) \	& 2.40 			& 2.53 & 2.40 & 2.53 \\
		\hline
		\vspace*{0.1cm}
		$d_{\mathrm{ X-X}}$ (\AA) \  & 3.13 			& 3.53 & 3.15 & 3.37 \\
		\hline
		\vspace*{0.1cm}
		$\delta_{\mathrm{ X-M-X}}$ ($^{\circ}$)	&81.34	&82.90	&81.83	&83.29 \\
		\hline
		\vspace*{0.1cm}
		$d$ (\AA)	&6.16	&6.62	&6.23	&6.53 \\
		\hline
		\vspace*{0.1cm}
  	$t_{\mathrm{ eff}}$ (\AA)	&13.42	&14.37	&13.81	&14.25 \\
		\hline
	\end{tabular}
	\label{tab:latparam}
\end{table}

\textit{Methods.} Our first-principles calculations were performed using Quantum Espresso \cite{giannozzi2009quantum,qe2} simulation package. The electron-ion interactions were described via the projector augmented wave method \cite{PAW1,PAW2}, and the wave functions were expanded in a plane wave basis set with the cutoff of 800 eV. Based on the experimental structure parameters of bulk MX$_{2}$ \cite{MX2_stacking}, we constructed slabs consisting of two MX$_{2}$ layers. A vacuum region of 20 \AA \ along the $c$ axis was used to separate the spurious images of bilayers to minimize their interactions. The internal degrees of freedom were further relaxed until the forces were below 10$^{-3}$ Ry Bohr$^{-1}$, employing the Perdew-Burke-Ernzerhof (PBE) parametrization of generalized gradient approximation and vdW corrections (DFT-D3) \cite{PBE_GGA}. BZs were sampled on $k$ meshes of 12$\times$12$\times$1. The optimized structural parameters are given in Table \ref{tab:latparam}. The electric polarization was calculated using the Berry phase approach implemented in Quantum Espresso. However, it is important to mention that the accurate procedure requires the calculation of the centrosymmetric phase, which is not applicable in the case of parallel-stacked MX$_2$ bilayers. To obtain the values of $P_z$ and $P_y$, we substracted the values $P^0_z$ and $P^0_y$ calculated in the stacking configurations, where these components, instead of the full polarization vector, should vanish by symmetry. Spin-orbit coupling was taken into account self-consistently in all the simulations except for the relaxations and Berry phase calculations. The post-processing calculations of band structures and spin textures on dense $k$-meshes were performed using \textsc{paoflow} code \cite{paoflow1, paoflow2}.\\

\begin{acknowledgments}
We thank Bertjan van Dijk for helpful contribution. J.S. acknowledges the Rosalind Franklin Fellowship from the University of Groningen. J.S. and E.B. acknowledge the funding from NWO under the contract NWA.1418.22.014. Part of the work was supported by the National Science Centre, Poland, under the Grant No, 2018/30/E/ST5/00667. The calculations were carried out on the Dutch national e-infrastructure with the support of SURF Cooperative (EINF-5312) and on the H\'{a}br\'{o}k high-performance computing cluster of the University of Groningen.
\end{acknowledgments}

\bibliography{bibfile}

\end{document}